\documentstyle[12pt]{article}
\def\be{\begin{equation}}
\def\ee{\end{equation}}
\begin{document}

\begin{center}

{\bf Half-Integral Spin-Singlet Quantum Hall Effect}

\vspace{1cm}

L. Belkhir, X.G. Wu, and J.K. Jain

\vspace{1cm}

{\em Department of Physics, State University of New York, Stony
Brook, New York, 11794-3800}

\end{center}

\vspace{2cm}

\noindent

We provide numerical evidence that the ground state of
a short range interaction model at $\nu=1/2$
is incompressible and spin-singlet for a wide range of repulsive
interactions. Furthermore it is accurately described by a trial
wave function studied earlier.
For the Coulomb interaction we find that this wave function provides
a good description of the lowest lying spin-singlet state, and
propose that fractional quantum Hall effect would occur at
$\nu=1/2$ if this state became the global ground state.
Some conclusions in an earlier paper are invalidated by the present
numerical study of a larger system; these will be indicated.

\pagebreak

A large number of fractions have been observed in the fractional
quantum Hall effect (FQHE) in a single two-dimensional layer of
electrons \cite{tsg,pg}.  With the
exception of 5/2 \cite{willet}, all have odd-denominators
\cite{note1}.
An investigation of the 5/2 FQHE is the objective of this paper.
Let us start by enumerating several relevant facts.
(i) FQHE has not been observed at the related filling factor
$\nu=1/2$. At $\nu=5/2$ the
FQHE is seen in a relatively small range of parameters.
(ii) Tilted field experiments have shown that the 5/2 FQHE is quickly
destroyed by raising the Zeeman energy \cite{eisen}.
This suggests that the
incompressible state is not fully polarized. We will assume, following
Haldane and Rezayi (HR) \cite{hr}, that it is spin-singlet. (iii) In
numerical
calculations investigating the nature of the states at $\nu=1/2$ and
5/2, pure two-dimensional Coulomb interaction does not produce a
spin-singlet ground state.
In fact, there is good numerical evidence that the thermodynamic ground
state at $\nu=1/2$ is fully polarized even for vanishing Zeeman
energy \cite{morf}.
We wish here to point out here that our claim in an earlier paper
\cite{bj}, based on a study of systems of
4 and 6 electrons, that the Coulomb ground state is spin-singlet at
$\nu=1/2$ has been invalidated by the present 8 electron
calculation \cite{note2}.

Even though FQHE has not been observed (yet) at $\nu=1/2$ in
single-layer systems, in
numerical calculations it is convenient
to work at $\nu=1/2$ rather than at $\nu=5/2$.  Following HR,
we will assume that the physics of the FQHE at
5/2 can be investigated by replacing the Coulomb
interaction at 1/2 by a model interaction which
simulates the conditions at $\nu=5/2$.
Another advantage of working within the lowest LL is that one can use
Haldane's pseudopotentials $V_{m}$ \cite{hal1}, which completely
specify the interaction. $V_{m}$ is the interaction energy of two
electrons in a state with relative angular momentum $m$.

The FQHE at 5/2, and it potential observation at 1/2 in single layer,
is still exciting a great deal of
interest, and so far no consensus has emerged around a unique
theoretical explanation. Haldane and Rezayi initially
proposed a `pairing' mechanism in the context of the hollow core
model that gives rise to a spin-singlet incompressible state
at $\nu=5/2$, described by a hollow core state. However the hollow
core state requires a considerably reduced repulsive interaction
at short range between the electrons in order to be stablized.
More recently, there has been
interest in the nature of the {\sl compressible} state at $\nu = 1/2$
as Halperin {\sl et al} suggested \cite{hlr}, in the spirit of the
composite fermion theory of the FQHE \cite{jain1},
that the
filling factor 1/2 corresponds to a Fermi liquid of spin
polarized composite fermions (electrons that cary two flux quanta).
The present work is, however, related to the possibility of an
incompressible state at 1/2.

In this paper we propose  the following scenario for the half
integral FQHE.
First we adopt an idealized short range interaction model
characterized by the following choice of Haldane pseudopotentials
$V_{m}=[\alpha,1,0,0,...]$. We find that the lowest energy
spin-singlet (LESS) state of the 8 electron system shows all the
properties
of an incompressible state. This state remains the global ground state
of the system up to $\alpha = 3.1$, beyond which a level
crossing
occurs, and the LESS state becomes an excited state.
(The global ground state for $\alpha > 3.1$ is not incompressible).
Furthermore we show that the
LESS state is accurately described by a either the Haldane-Rezayi
state ($\alpha \leq 1$)  or a trial wave function proposed
earlier by one of us \cite{jain2} (for $\alpha \geq 1$) .
For the Coulomb interaction, the LESS state is still well described by the
trial wave
function, but the global ground state is not spin-singlet or
incompressible.
We expect that FQHE will occur at $\nu=1/2$ if the LESS becomes the
global ground state. What (if anything) will make the singlet state the
overall ground state is not completely clear at the moment, but
we argue that LL mixing can possibly lower the energy of
this state sufficiently to make it the ground state.

Our trial wave function is given by
\be
\chi_{_{1/2}}=[\prod_{j<k=1}^{N}(z_{j}-z_{k})]\;
[\prod_{j<k=1}^{N/2}(z_{j}-z_{k})]\;[\prod_{j<k=\frac{N}{2}+1}
^{N}(z_{j}-z_{k})]\;
\chi_{_{2}}
\ee
where $z_{j}=x_{j}-iy_{j}$ denotes the position of the $j$th electron,
$z_{1},...,z_{N/2}$ refer to spin-up electrons,
$z_{N/2+1},...,z_{N}$ refer to spin-down electrons, and $\chi_{_{2}}$ is
the wave function of the state with two filled Landau levels,
constructed as though the electrons were spinless.
$\chi_{_{1/2}}$ is a singlet state because
it is given by a completely symmetric factor (symmetric with respect
to the exchange of any two coordinates) times the spin-singlet state
$$[\prod_{j<k=1}^{N/2}(z_{j}-z_{k})]\;[\prod_{j<k=\frac{N}{2}+1}
^{N}(z_{j}-z_{k})].$$ (This state clearly is singlet because
it corresponds to fully occupied spin-up
and spin-down states in the lowest LL.)

The appeal of this trial wave function becomes clear by noting
the remarkable fact that {\em all}
observed odd-denominator incompressible states have the structure
\cite{jain1}
\be
\chi_{_{n/(pn\pm 1)}}=\prod_{j<k=1}^{N}(z_{j}-z_{k})^{p}
\chi_{{\pm n}}\;\;,
\ee
where $\chi_{{\pm n}}$ is the wave function of IQHE state
att $\nu=n$ (with magnetic
field pointing in the $\pm z$ direction), and $p$ is an even integer.
In other words, other than the fact that each electron has captured
$p$ vortices, FQHE states are the same as the IQHE states. The bound state
of an electron and $p$ vortices can be interpreted as a particle,
called `composite fermion' (CF), and the  FQHE of electrons can be
interpreted as the IQHE of CFs.
In the limit of $B\rightarrow\infty$, when the FQHE states are maximally
polarized, electrons are
taken to be spinless in the construction of $\chi_{{\pm n}}$. In
the limit when the Zeeman energy is negligible, (but $\hbar \omega_c$
is still large)
$\chi_{{\pm n}}$ contains $n_{1\uparrow}$ spin-up
and $n_{1\downarrow}$ spin-down (where $n=n_{1\uparrow}+n_{1\downarrow}$)
Landau bands occupied. For even $n$, $n_{1\uparrow}=
n_{1\downarrow}=n/2$ and the low-field state is spin-singlet. For odd
$n$ it has a non-zero spin.
The CF states have been tested numerically for a large number of
cases in both  limits, and found to be extremely good
representations of the actual Coulomb ground states \cite{dj,wdj}.
Furthermore, a straightforward
generalization of the CF wave functions provides a complete and
microscopically accurate description of the entire low-energy Hilbert
space of states at arbitrary filling factors \cite{dj}.

The wave functions of the type in Eq.(2) can be written only for
odd-denominator filling factors.
The spin-singlet wave function $\chi_{_{1/2}}$ is the simplest
generalization of these wave functions to an even-denominator
fraction.  It also
lends itself to a CF interpretation, since it contains vortices bound
to electrons in the state
$\chi_{_{2}}$. The difference from the states of
Eq.(2) is that now
two kinds of vortices are attached to each electron of $\chi_{_{2}}$:
one is seen by {\em all} other electrons, while the other is seen only
by electrons of the same spin. Note that it
is the electron spin that allows us to write a CF wave function at a
half-integral filling factor, which brings out the important role of
spin in the case of even-denominator FQHE.

Motivated by the success of the CF theory, we study
$\chi_{_{1/2}}$ in this paper using finite size exact
diagonalization techniques. Since we are interested in a singlet
state, we set the Zeeman energy to zero in all our calculations.
We start with a short-range interaction model , defined by the
pseudopotential parameters
\be
[V_{0},V_{1},V_{2},...]=[\alpha,1,0,0,...]\;\;.
\ee
The only variable in this model is
$\alpha=V_{0}/V_{1}$.
Short-range-interaction models have proved quite successful in reproducing
the phenomenology of the FQHE \cite{hal2}; for example, the
model  of eq.(3)
produces FQHE at $n/(2n\pm 1)$ for fully polarized electrons
\cite{gmcd}.  The reason is that incompressible states are not
very sensitive to the details of interaction. We will show that
at $\nu = 1/2$ the LESS state of this model possesses the properties of
an incompressible state.
Another advantage of this model is that
its ground state is known exactly for $\alpha=0$, where it is precisely
given by the HR wave function. In numerical studies, it is found that
the HR ground state is valid roughly for $\alpha<1.0$, i.e., for
interactions that have an attractive core. We will now show that the
LESS state of this model for $\alpha > 1.0$ is well described by the
lowest LL projection of the CF trial wave function $\chi_{1/2}$.

We investigate the nature of the LESS state numerically.
Our numerical calculations are performed in the spherical
geometry \cite{hal1}, in which $N$ electrons move on the surface
of a sphere under the
influence of a radial magnetic field produced by a magnetic monopole
at the center. The flux through the sphere is given by $N_{\phi}hc/e$
where $N_{\phi}$ is an integer due to Dirac quantization
condition. The state $\chi_{_{1/2}}$ occurs when
\be
N_{\phi}=2N-4\;.
\ee
Clearly, in the limit of large $N$, the filling factor $N/N_{\phi}$ is
1/2. We work with eight electrons. The total size of the Hilbert space
in the lowest LL is 1,562,275. However, due to
the symmetry of the problem, it is sufficient to
work in the sector with $L_{z}=S_{z}=0$, where $L_{z}$ and $S_{z}$ are
the $z$ components of the total orbital angular momentum and the total
spin, respectively. In this sector, the size of the Hilbert space is
21,773, which is numerically manageable by Lanczos techniques.
For an eight electron system, we find, coming from above, that there
is a transition from a non-singlet ground state
to a singlet ground state at $\alpha=3.2$, and the ground state remains
singlet for $\alpha<3.2$.

We compare the LESS state of the above model (which is also {\em the}
ground state for $\alpha<3.2$) with
three trial wave functions. One is HR hollow-core
wave function $\chi_{1/2}^{HR}$, which
is the exact ground state of the above one parameter model when
$\alpha=0$. (This wave function also occurs at $N_{\phi}=2N-4$).
The second trial wave functions is the lowest LL
projection of our trial wave function $\chi_{_{1/2}}$:
\be
{\cal P}\chi_{_{1/2}}={\cal P}\;[\prod_{j<k=1}^{N}(z_{j}-z_{k})]\;
[\prod_{j<k=1}^{N/2}(z_{j}-z_{k})]\;[\prod_{j<k=\frac{N}{2}+1}
^{N}(z_{j}-z_{k})]\;
\chi_{_{2}}\;,
\ee
where ${\cal P}$ is the projection operator. Unlike $\chi_{_{1/2}}$,
this projected wave function does not vanish when two electrons of
opposite spin approach one another, and is not expected to be very
accurate in the limit of $\alpha\rightarrow\infty$. To redress this
problem, we construct the following `hard-core' trial wave function:
\be
{\cal P}_{\infty}\chi_{_{1/2}}=[\prod_{j<k=1}^{N}(z_{j}-z_{k})]\;{\cal P}\;
[\prod_{j<k=1}^{N/2}(z_{j}-z_{k})]\;[\prod_{j<k=\frac{N}{2}+1}
^{N}(z_{j}-z_{k})]\; \chi_{_{2}}
\ee
This explicitly vanishes when any two electrons coincide, regardless
of their spin, and is
expected to be more appropriate in the limit of
$\alpha\rightarrow\infty$.
The projection is carried out using techniques described elsewhere
\cite{dj}.

The overlap between the hollow-core and the hard-core trial wave
functions is quite small (0.0377), which is not surprising given their
distinct physical origins. For $\alpha<<1.1$, the LESS state is
expected to be close to the HR hollow-core wave function.
Our numerical calculations show that for $\alpha>>1.1$,
the hard-core trial wave function
is indeed a good representation of the LESS
state. [For example, the overlap between
the true LESS state with the hard-core trial wave function  is 0.96
for $\alpha=10$.] Thus, we have a good understanding
of the two extreme limits of the one-parameter short range model.
Fig.1 shows how the various overlaps vary as a function of $\alpha$
in the regime where the LESS state is the global ground state.
The hard-core state does better than the hollow-core
state for $\alpha\geq 1.1$. This is the region where the
interaction at short distances is repulsive; $\alpha<1.1$ effectively
corresponds to an attractive core.
Interestingly, in the intermidiate regime, $0.7<\alpha<2.7$, the ground
state is best described by the simply projected state ${\cal
P}\chi_{_{1/2}}$. Thus, {\em a remarkably good description of
the short-ranged model is possible in the entire parameter regime}.
The fact that the LESS state is well approximated by these trial wave
functions also implies that it is incompressible, as these trial wave
functions describe incompressible states.

Unfortunately, we do not know of a realistic model
which exhibits singlet FQHE at $\nu=1/2$.
The following calculations provide some insight into the
relevance of our results to the real case with the Coulomb interaction.
First of all, the LESS state of the Coulomb interaction
(which is not the ground state) is quite close to the CF trial
wave function (overlap of 0.82 with the hard-core wave function). The
question is if it can be made to be the ground state. In
Fig. 2, we show the behavior of the ground state for an
interaction $[\alpha V_{1}, V_{1},V_{2}, ...]$,
in which all $V_{i}$, are set at their Coulomb
values (at $\nu=1/2$), which we will call `modified Coulomb
interaction'.
The physical Coulomb interaction corresponds to $\alpha=2.0$.
Here also we see that the global ground state is spin-singlet for up
to $\alpha=1.4$, at value a crossover occurs a ground state which is
compressible ($L=1$), and the LESS state
becomes an excited state. Similar to the short-range model,
the LESS state is best decribed by the hollow core state for $\alpha<1.1$,
whereas for $\alpha>1.1$ it is quite well described by the CF states.

Thus, if by tuning some parameters, the singlet state could be made
the ground state, FQHE would result at $\nu=1/2$.  What will make
the ground state singlet? It is possible that there will be
a level crossing transition to a singlet ground
state as LL mixing is increased. Unfortunately, calculations
with 4 or 6 electrons are not reliable because of finite size
effects, while for an eight particle system, it is not possible
for us to carry out a direct numirical investigation
of the effect of LL mixing due to the enormously large Hilbert space.
However, the following points support our belief. (i) As shown
by Rezayi and Haldane \cite{rh}, LL mixing effectively renormalizes the
pseudopotentials $V_{m}$ in such a way that $V_{0}$ is reduced
more rapidly than the others. Consider a situation in which all
$V_{i}$, $i\neq 0$, are held fixed at their Coulomb values,
while the `contact' pseudopotential, $V_{0}$, is reduced. Clearly,
this does not affect the energy of the fully polarized state, but
reduces the energy of the singlet eigenstates. Therefore, it is
plausible that if the contact interaction
is reduced sufficiently, the singlet state may become
the ground state \cite{hrnote}.
This is explicitly seen to be true in the extreme
limit when $V_{m}=[0,1,0,0,...]$.
In this case, there is a unique zero energy singlet ground state,
given exactly by HR trial wave function \cite{hr}.
(ii) Now let us consider the other extreme limit of infinite
hard-core repulsion, when the pseudopotential parameters are given by
$V_{m}=[\infty,1,0,0,...]$. The ground state in the lowest LL is not
singlet for this model. Let us now include two LLs, and vary the LL
separation to govern the LL mixing \cite{note3}.
When the LL spacing is zero, it can be proven that there is again a
unique zero energy singlet ground state given by
our trial wave function $\chi_{_{1/2}}$ (see Appendix).
Thus, even for very large $V_{0}$, LL mixing can produce a
singlet ground state. This is a rather remarkable result, since,
in general, strong short-range repulsion favors fully polarized states.
We also find that in the two LL model the hard core interaction mimicks
the Coulomb interaction quite faithfully. In our 4 electron
calculation, the Coulomb ground state with zero LL spacing has an
overlap of 0.96 with $\chi_{1/2}$. Unfortunately, for 6 and 8
electrons the Hilbert space of the 2 LL problem is prohibitively
large, which prevents us  from studying the effects of LL mixing.
But, clearly, as the LL mixing is increased by reducing
the LL spacing, there is an explicit level crossing transition to a
singlet ground state in this model.

We parenthetically note here that we had claimed in Ref[8] that in the
two LL model the ground state of the hard-core Hamiltonian of Eq.(3)
at $\hbar\omega_c = 0$ (which is our trial wave function) is
adiabatically connected to the ground state at $\hbar\omega_c = \infty$.
Clearly, this is not correct, since the ground state at $\hbar\omega_c
= \infty$ is not even spin-singlet. Our calculation show, however,
that the $\hbar\omega_c = 0$ ground state evolves adiabatically into
the LESS state at $\hbar\omega_c  = \infty$.

We have also investigated the finite width effects on the ground state
of the system. We looked at the simplest case of a square well
potential, as well as the case where there is a small potential
barrier in the center of the well, which leads to an effective double
layer system \cite{gww}. It was found that finite width does not
stabilize the
spin-singlet state in either case.
Even though finite width gives the desired effect of reducing the ratio
$V_0/V_1$, it also makes the effective interaction more long-ranged,
which in turns lowers the critical ratio of  $V_0/V_1$ at
wich the crossover to a spin-singlet state occurs.

In conclusion, we have proposed the following scenario for the FQHE at
$\nu=1/2$. The lowest energy state {\em in the spin-singlet sector} is
`incompressible', and is well
represented by a CF trial wave function for repulsive interactions.
However, it is not the global ground state for most parameters,
which is the reason why FQHE at
half-integral filling factor is not observed.
We propose that half-integral FQHE
will occur for those parameters for which the singlet
state becomes the overall ground state.
This is the main result of our work.

Now we investigate the relevance of $\chi_{1/2}$ to the FQHE at
$\nu=5/2$. As mentioned earlier, we still work at $\nu=1/2$ (i.e in
the lowest LL), but choose the pseudopotentials appropriate for
$\nu=5/2$.
In this case the physical interaction corresponds to $\alpha=1.45$.
Fig. 3 shows the behavior of the LESS state.
Unfortunately, the CF states, even though showing the same pattern
as for
1/2, have quite a poor overlap with the LESS state. This is mainly due
to the relatively large value of $V_2$ in the second LL.
We have checked that a ratio $V_2/V_1 \leq 0.8$ is needed to make
the CF states relevant. The Coulomb ratios of $V_2/V_1$ are $0.766$
at 1/2 and $1.083$ at 5/2. This also explains why Fig.3 is so
different from the short-range model results.
At 5/2 the crossover  from a spin-singlet to a spin-polarized
ground state (not shown in Fig.3) occurs at $\alpha=1.2$,
Note however that the HR state shows quite a good overlap with the
ground state up to $\alpha=1.1$. The level crossing occurs relatively
much closer to
the Coulomb value in 5/2 than in 1/2, making it easier for LL mixing
(or some other mechanism) to produce incompressibility. Unlike at 1/2,
the incompressible state at 5/2 is likely to be better described by the
HR state than the CF states. Note also that, rather unexpectedly,
the HR state has a very small overlap at $\alpha=0$.

We thank Profs. B.I. Halperin, S. He, R. Morf, and X.C. Xie for
several useful discussions and communications, and especially Prof.
Morf for sending us the results of his unpublished calculations.
This work has been supported in part by the National Science Foundation
under Grant no. DMR-9020637.

{\bf APPENDIX}

Let us give the proof of the following theorem.

{\bf Theorem:} In the 2LL Hilbert space of $0\uparrow$,$0\downarrow$,
$1\uparrow$,$1\downarrow$ Landau bands, assuming all of them to be
degenerate at zero energy, the spin-singlet state $\chi_{1/2}$
is the unique zero energy ground state at $\nu = 1/2$ of the model
interaction
$$V_{TK}(r) = \infty~\delta(r) + \lambda\nabla^2\delta(r),$$
which is  equivalent to the infinite hard-core repulsion defined in
the text.

{\bf Proof:}All statets with zero energy must vanish at least as fast
as $r^2$ as two electrons, at a distance $r$, approach one another. We
now show that at $\nu = 1/2$, $\chi_{1/2}$ is the only state with this
property. (i) Since only two LLs are available, one of the zeros
as the $j$ and $k$ electrons are brought close to each other must be
of the form $(z_j - z_k)$. Thus the wave function must contain a
factor $\chi_1 = \Pi_{j<k=1}^{N}(z_j - z_k)$.\\
(ii) Due to Pauli principle, when two electrons with the same spin
approach each other, the wave function must vanish at least as fast as
$r^3$. Thus the state must contain another factor
$\chi_{1;1} = \Pi_{r<s=1}^{N/2}(z_r - z_s)~\Pi_{p<q=N/2+1}^{N}(z_p -
z_q)$.\\
(iii) Thus the most general form of a state that is confined to the
lowest LLs and vanishes at least as fast as $r^2$ is
$$ \chi_{1/2} = \chi_1\chi_{1;1}\chi_{\nu}$$
where $\chi_{\nu}$ must be within the lowest two LLs. Further note
that $\chi_{\nu}$ must vanish when {\sl any} two electrons coincide;
for electrons with opposite spins it must vanish because we want the
wave function to vanish at least as fast as $r^2$ as any two
electrons come close, and for electrons with the same spins it must
vanish due to Pauli principle. The largest value that $\nu$ can assume
is $\nu = 2$ where $\chi_{\nu}$ is the state with two filled LLs of
spinless electrons. The product $\chi_{1/2}$ is then the unique
wave function at 1/2 which vanishes at least as fast as $r^2$ when two
electrons approach each other. It is therefore the unique zero energy
ground state for the above hard-core model interaction.

\pagebreak


\pagebreak

{\bf Figure Captions}

Fig.1. Overlap of the exact ground state of the short range model
defined in the text with the hollow-core state $\chi_{_{1/2}}^{HR}$
(dashed line), the hard-core state ${\cal P}_{\infty}\chi_{_{1/2}}$
(dotted line), and with the simply projected state ${\cal
P}\chi_{_{1/2}}$ (solid line). The calculation in this figure as well
as those in Figs. 2 and 3 was done for an 8 electron system, and at zero
Zeeman energy.

Fig.2. Overlap of the exact ground state of the modified Coulomb
model in the lowest LL ($\nu = 1/2$)
with the hollow-core state $\chi_{_{1/2}}^{HR}$
(dashed line), the hard-core state ${\cal P}_{\infty}\chi_{_{1/2}}$
(dotted line), and with the simply projected state ${\cal
P}\chi_{_{1/2}}$ (solid line).
The actual Coulomb interaction occurs at $\alpha=2.0$.

Fig.3. Overlap of the lowest energy spin-singlet eigenstate
of the modified Coulomb model in
the first LL ($\nu = 5/2$)
with the hollow-core state $\chi_{_{1/2}}^{HR}$
(dashed line), the hard-core state ${\cal P}_{\infty}\chi_{_{1/2}}$
(dotted line), and with the simply projected state ${\cal
P}\chi_{_{1/2}}$ (solid line). This state is the ground state for
$V_0/V_1 < 1.2$. The actual Coulomb interaction occurs at $\alpha=1.45$.


\begin{thebibliography}{99}
\bibitem{note2} For 8 electrons, the Coulomb ground state at $\nu=1/2$ has
$S=1$. We believe based on Morf's calculations\cite{morf}, that the ground
state
will be fully polarized as the system size is increased.
\bibitem{hrnote} Haldane and Rezayi
investigated if Landau level mixing could be responsible for such
reduction, but found that not to be the case in their perturbative
calculation.
\end{thebibliography}
\end{document}